# Optically Levitated Nanoparticles as Receiving Antennas for Low Frequency Wireless Communication

Zhenhai Fu[1,4], Jinsheng Xu[1], Shaochong Zhu[1], Chaoxiong He[1], Xunming Zhu[1], Xiaowen Gao[1,5], Han Cai[2], Peitong He[1], Zhiming Chen[1], Yizhou Zhang[1], Nan Li[2], Xingfan Chen[2], Ying Dong[1], Shiyao Zhu[2,3], Cheng Liu[2], Huizhu Hu[1,2,6]

[1]*Quantum Sensing Center, Zhejiang Lab, Hangzhou 310000, China*
[2]*State Key Laboratory for Extreme Photonics and Instrumentation, College of Optical Science and Engineering, Zhejiang University, Hangzhou 310027, China*
[3]*Zhejiang Province Key Laboratory of Quantum Technology and Device, School of Physics, Zhejiang University, Hangzhou 310027, China*
[4]*fuzhenhai@zju.edu.cn*
[5]*gaoxw@zhejianglab.com*
[6]*huhuizhu2000@zju.edu.cn*

**Abstract:** Low-frequency (LF) wireless communications play a crucial role in ensuring anti-interference, long-range, and efficient communication across various environments. However, in conventional LF communication systems, their antenna size is required to be inversely proportional to the wavelength, so that their mobility and flexibility are greatly limited. Here we introduce a novel prototype of LF receiving antennas based on optically levitated nanoparticles, which overcomes the size-frequency limitation to reduce the antenna size to the hundred-nanometer scale. These charged particles are extremely sensitive to external electric field as mechanical resonators, and their resonant frequencies are adjustable. The effectiveness of these antennas was experimentally demonstrated by using the frequency shift keying (2FSK) modulation scheme. The experimental results indicate a correlation between error rate and factors such as transmission rate, signal strength, and vacuum degree with a signal strength of approximately 0.1V/m and a bit error rate below 0.1%. This advancement in leveraging levitated particle mechanical resonators (LPMRs) as LF antennas marks a significant stride in long-distance communication technology.

**Introduction.** Wireless communications in low-frequency band have been widely investigated and have various applications, such as underwater, underground, and earth-ionosphere waveguide because of its long propagation distance, strong penetration capability, and resistance to interference [1]. The prevalence of portable or autonomous platforms in these applications has significantly increased the demand for smaller, more efficient and higher sensitive transmitting and receiving antennas [2]. Conventional LF antennas in the electrically small regime, such as magnetic induction coils and electric dipole antennas, face fundamental limitations in terms of size-efficiency trade-offs when they are used as transmitting antennas, and size-sensitivity trade-offs when they are used as receiving antennas [3]. The primary solution for miniaturizing LF transmission antennas involves mechanical antennas based on permanent magnets [4,5], electrets [6,7], and piezoelectric materials [8,9,10,11]. Their radiation principles eschew the dependence on electronic circuit oscillating currents for radiation generation and, instead, harness mechanical energy to drive the motion of charges or magnetic dipoles, thereby breaking the size-efficiency constraints imposed by the wavelength for the LF transmission antennas [1]. Consequently, miniaturizing receiving antennas becomes a major challenge and research hotspot in the LF communication systems.

To address the miniaturization of receiving antennas, most of the previous research was focused on magnetoelectric coupling antennas [12]. However, when the magnetoelectric coupling antennas are used as receiving antennas, their miniaturization is still constrained by the matching of signal wavelength and antenna physical dimensions. The resonant frequency $f$ is inversely proportional to the antenna size ($f \propto L^{-1}$ [13]) and is influenced by material properties of the piezoelectric and magneto strictive materials [14]. Therefore, the size of magnetoelectric antennas in the LF band are restricted to centimeter scale. Moreover, to enhance receiving sensitivity and achieve frequency tunability, an external magnetic biasing, such as permanent magnets [15,16,17,18], is required — an essential configuration for most magnetic sensors like Nitrogen-vacancy (NV) center in diamond [19,20].

Recently, vacuum-levitated nano-micro particles, characterized by their non-mechanical dissipation and low thermal dissipation, have garnered widespread attention due to their high sensitivity in detecting external forces [21,22,23,24], torques [25,26,27], accelerations [28,29], and masses [30,31,32]. Our research team has previously demonstrated that optically levitated particle mechanical resonators (LPMRs) can be developed into high-sensitivity electric field nano-sensors [33]. This development relies on achieving ultra-high force detection sensitivity [24] in the vacuum and enabling them to carry a high net charge. The LPMRs appear to be inherently tailored for the LF communication antennas due to their wavelength-independent resonance frequency and tunability. On one hand, the resonant frequency $f$ of LPMRs naturally falls within the LF band, which is primarily determined by parameters of optical potential well and independent of the signal wavelength and physical dimensions of particles as antennas [34]. This fundamentally breaks the size-frequency constraints on the antennas, reducing characteristic dimensions for the LF antennas from the centimeter scale to the hundred-nanometer scale (nearly five orders of magnitude). On the other hand, in terms of receiving mechanisms, the minimum detectable field of traditional dipole antennas is inversely proportional to the antenna's physical length ($E_{min} \propto L^{-1}$) and depends on impedance matching between the antenna and the load resistance [35]. In contrast, the sensitivity of LPMRs is determined by its force detection sensitivity, directly proportional to the three-halves power of the particle's size ($E_{min} \propto L^{3/2}$) and can be enhanced by the net charge $nq_e$ carried by the LPMRs [33]. Such a relationship between size and sensitivity ensures that reducing size and improving sensitivity are pursued as parallel objectives rather than as a trade-off. Moreover, the receiving bandwidth of traditional antennas is similarly constrained by impedance matching between the antenna and the load circuit, which is challenging to balance tunability and high sensitivity. However, the LPMRs are not subject to such constraints. Their tunability [36] and capability to be arrayed [37,38] allow for a significant enhancement in the overall bandwidth and carrier spectrum range for communication applications. Finally, the vector-detection capability of the LPMRs makes them applicable to a broader range of communication modulation schemes and make the LPMRs potential omni-directional receiving antennas. This feature, compared to scalar-detection electromagnetic sensors, can offer advantages in communication scenarios.

In this study, we propose a novel method that utilizes levitated nanoparticle as receiving antenna to achieve ultra-sensitive LF communication, which can fundamentally break the size-sensitivity limitations of traditional antennas as receivers. The feasibility of our method is theoretically studied by using the frequency shift keying (2FSK) modulation scheme in digital communication, where we establish a model for communication error rates. The model is validated by experimental results demonstrating the relationship between the error rate and the transmission rate, signal strength, and vacuum degree. To demonstrate the optimal communication performance of the system, we pumped to high vacuums of $2\times10^{-7}$ mbar while tuning the resonant frequency by adjusting the trapping optical power. We achieve competitive LF communication performance in the order of 0.1V/m in the LF band, with an error rate lower than 0.1%. Compared to the existing LF receiving antennas, our method reduces the size constraint on the LF antennas from the centimeter level to the hundred-nanometer scale, which is nearly five orders of magnitude smaller. Our study can provide a suitable LF receiving antennas for long-distance communication.

**Experimental setup.** The experimental setup and the coordinate system are shown in Fig.1(a). A charged nanoparticle, levitated within a stable optical trap, acts as a sensitive antenna. The optical trap is created by focusing a collimated laser beam with a high convergent angle. The collimated laser beam is linearly polarized along the *x* axis and propagates along the *z* axis. This setup facilitates communication through the electric field, which carries the LF transmitted signal. See the 'Methods' section for more details. As shown in the inset of Fig.1(a), in the absence of electric forces, the trapped nanoparticle exhibits damped harmonic oscillations at distinct frequencies in three orthogonal directions around its equilibrium positions. Therefore, the levitated nanoparticle has the advantage to perform LF communication within a wide frequency range and the potential to develop frequency division multiplexing communications. Under the influence of an external electric field, the particle's oscillations are modulated accordingly. In the transmission of digital information, the widely-used 2FSK modulation scheme employs two distinct carrier frequencies. Symbols '0' and '1' correspond to carrier frequencies $f_1$ and $f_2$, respectively (as shown in Fig. 1b). The resulting communication signal can be viewed as the particle's motion signal, distorted by the transfer function of the resonator and superimposed with base noise, primarily from thermal and light source effects. Monitoring the scattered light to track the temporal motions of the particle allows us to demodulate the communication signal from the particle's motion signal. Further insights into the communication principles and the complete signal demodulation process can be found in Appendix A.

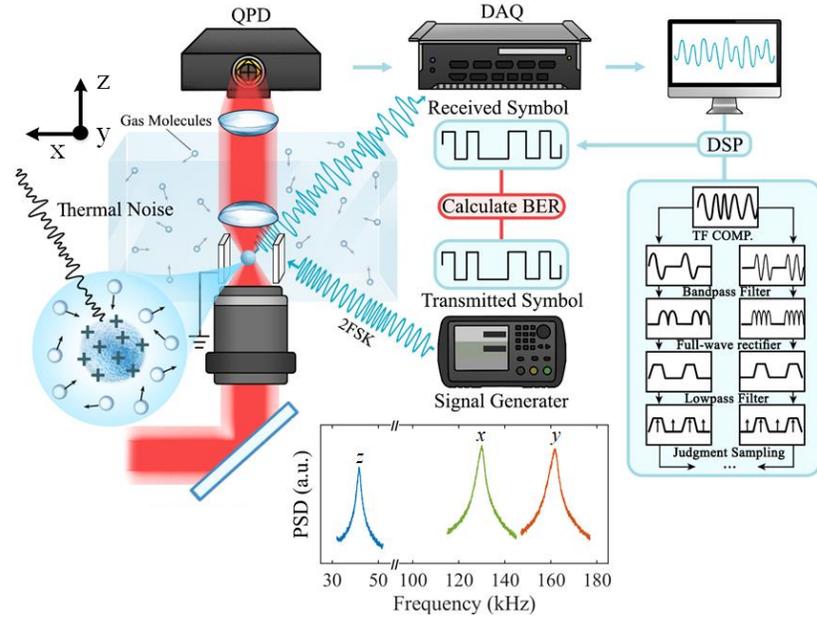

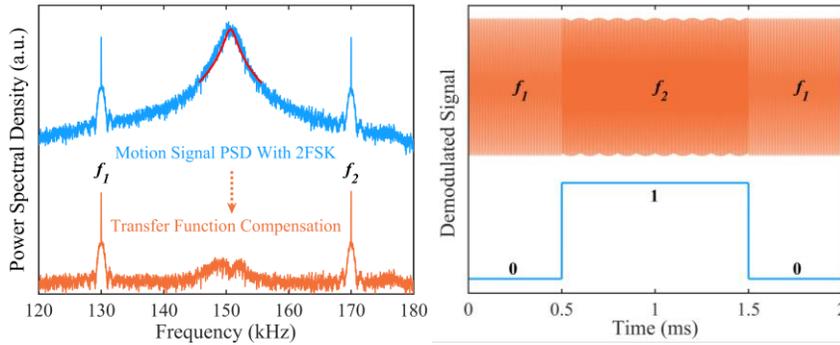

**Fig. 1.** LF communication. **a.** The experimental setup consists of a signal generator (for generating 2FSK code), a pair of needle electrodes (for generating LF communication signals), an optical trap (for levitating a charged nanoparticle within a vacuum chamber), a quadrant photoelectron detector (for detecting scattered light), a data acquisition system (DAQ), and digital signal processing (DSP). Further details are provided in the 'Methods' section. **b.** Power spectral density (PSD) showing the 2FSK communication signal at 1mbar (upper curve) and the compensated signal with transfer function (lower curve). **c.** Communication symbols: The two schematic carrier frequencies are 130kHz and 170kHz, with a signal transmission rate of 1kbit/s. The processed received signal, illustrated in the decision-based bit sequence plot below, yields the bit sequence through the DSP.

**Bit error rate (BER)**, a key performance metric for communication systems, is the ratio of incorrect symbols in received signals to the total transmitted symbols. Compensating for the resonator's transfer function mitigates its influence, leaving the communication system predominantly affected by white base noise. Therefore, the BER formula of 2FSK non-coherent demodulation is applicable. The transfer function of LPMR directly impacting the demodulated BER, determined by the mass $m$, the resonant frequency $\omega_0$, and damping rate $\Gamma_0$, as shown in Equation (1):

$$\chi(\omega) = m^{-1}\left[\left(\omega_0^2 - \omega^2\right) + j\omega\Gamma_0\right]^{-1} \quad (1)$$

Given that air pressure (affecting damping rate) significantly influences the transfer function, stabilizing the air pressure and collecting data without electric field drive for an extended period is crucial before each communication experiment to accurately determine the transfer function. We first acquired a time-domain signal with 65536 data points at a sampling rate of 937.5 kHz. The PSD near the resonance frequency was fitted with a Lorentzian curve, and the process was repeated 100 times to obtain the average damping rate and resonance frequency. Subsequently, we employed Eq.1 to derive the transfer function of LPMR for various experimental conditions.

Assuming a constant net charge for the particle and linear dependence of optical forces on spatial coordinates, we developed a theoretical model. It infers that our system's BER correlates with the signal amplitude $E_V$, the mass $m$, the signal

transmission rate $f_T$, and the damping rate $\Gamma_0$, as shown in Equation (2):

$$\text{BER} = \frac{1}{2}\exp\left[-\frac{(E_V Nq_e)^2}{32k_B Tm\Gamma_0 f_T}\right] \quad (2)$$

where $T$ is the equivalent center-of-mass temperature and $k_B$ is the Boltzmann constant. See Appendix B for detailed derivations.

**Results.** To characterize the 2FSK communication, we conduct communication experiments in the $x$-axis mode with a resonant frequency of $f_0 = 150$ kHz, selecting symmetrical carrier frequencies $f_1 = 130$ kHz and $f_2 = 170$ kHz around the resonant frequency. The net charge of the nanoparticle remains constant at $Nq_e = 3e$. We systematically vary the amplitude of the electric field signal, transmission rate, and pressure separately, obtaining the corresponding demodulated BER under different parameter values. Fig. 2 illustrate the experimental and theoretical BER values under various parameter cases.

The agreements between the experiments and the theoretical model validate the effectiveness of our communication system. As depicted in Fig.2a, both theoretical and experimental results indicate that the BER decreases with an increase in the signal amplitude. Below approximately 9 kV/m amplitude, experimental and theoretical results align, diverging at higher amplitudes. This discrepancy is due to LPMR's oscillation at large amplitudes in regions with significant nonlinearity in optical forces. In Fig.2b, experiments were conducted in an approximate pressure range from 1 mbar to 10 mbar. Experimental and theoretical results show nearly perfect agreement at low vacuum. Although inconsistencies at high vacuum, likely due to nonlinear effects, persist, both results affirm that high vacuum is essential for optimal BER. Equation (2) illustrates that the BER increases with an increase in the transmission rate. In Fig.2c, results align for transmission rates above 2 kbit/s, with lower rate inconsistencies likely due to the limited signal length.

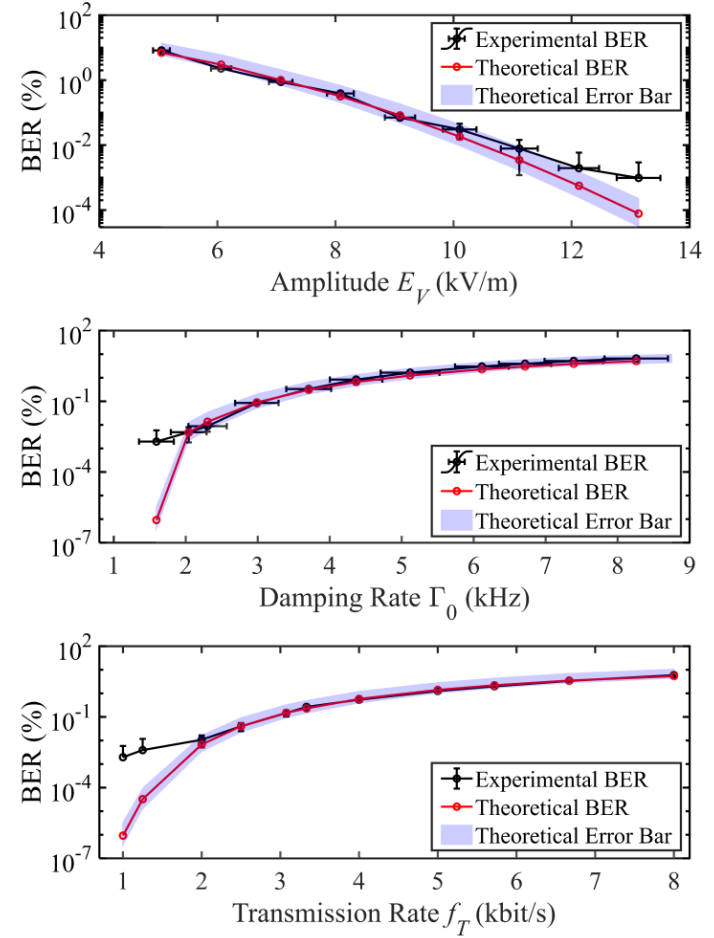

**Fig. 2.** Characterizing the BER of the LF communication system with a resonant frequency of around 150kHz and a constant net charge $Nq_e=3e$. (a) BER variation with signal amplitudes when the damping rate and transmission rate are approximately 1.59 kHz and 1 kbit/s, respectively. (b) BER variation with damping rates when the signal amplitude and the transmission rate are approximately (15.156±0.427) kV/m and 1 kbit/s, respectively; (c) BER variation with transmission rates when the damping rate and the signal amplitude are approximately 1.59 kHz and (15.156±0.427) kV/m, respectively. The red curve depicts the BER calculated with theoretical parameters, and the purple error bar area indicates the theoretical BER's error margin, factoring in parameters deviations of $E_V$, $m$, $T$, $\Gamma_0$ (see Appendix B for details). The black curve shows the experimental BER. For all cases, the transmitted signal, totaling 102350 bits, was segmented into 10 groups. For each case, the average BER and the standard deviation across the 10 groups are presented as the measured value and corresponding error bar, respectively.

The results clearly show that large signal amplitude, high vacuum degree, and low transmission rate significantly reduce the BER. However, reducing transmission rate implies an increase in the time required to complete a communication process, introducing a trade-off in certain application scenarios. Moreover, beyond BER, electric field detection sensitivity is another critical index for characterizing the performance of LF communication systems. Sensitivity, limited by thermal noise, necessitates ultrahigh vacuum to achieve optimal BER and sensitivity. However, achieving low BER with large signal amplitude contradicts the aim for optimal sensitivity, presenting a challenge for enhancing communication performance. Since the transmitted electric field modulates LPMR's oscillations, generating strong electric forces with weaker fields is key to overcoming this challenge. Therefore, imparting a higher net charge to the LPMR is an effective strategy for further optimizing the communication system. By irradiating the nanoparticles with a focused electron beam in vacuum, we induce and stabilize a high level of elementary charges on them. The amount of charge can reach more than 200, see the 'Methods' section for details.

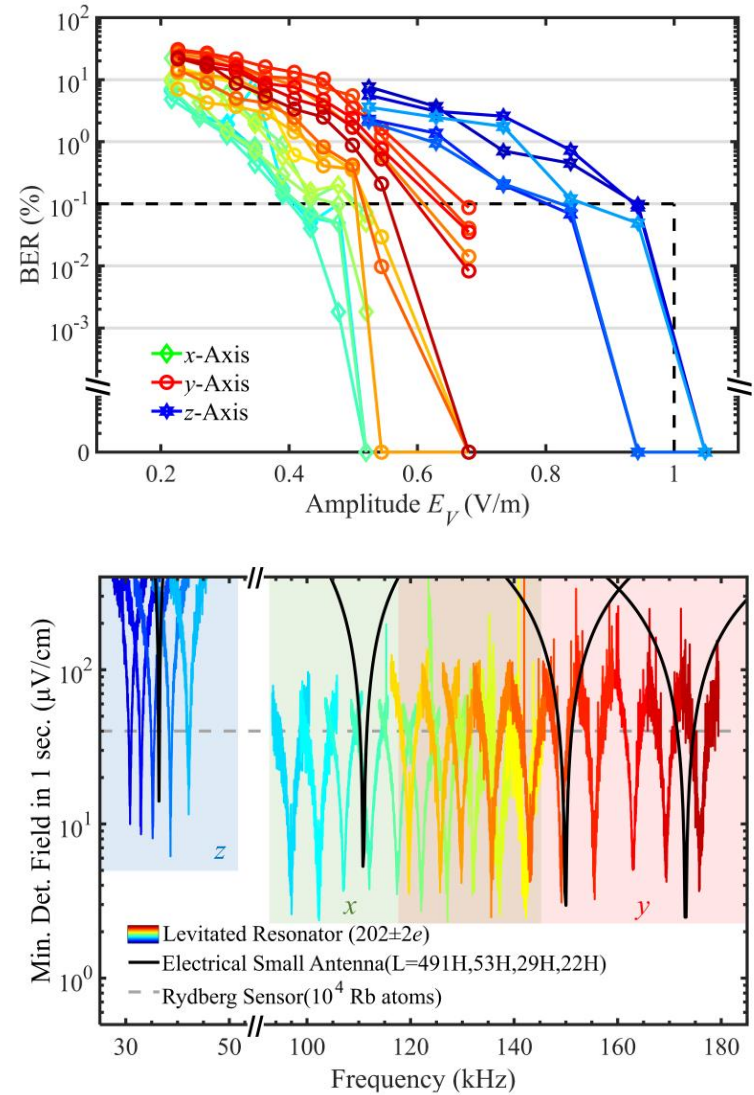

**Fig. 3.** Communication performances at $2\times10^{-7}$ mbar with a transmission rate of 0.5 kbit/s. (a) BER depending on signal amplitude $E_V$ at different resonant frequencies. (b) Minimum detectable electric field at different frequencies compared with the electric small antenna and the Rydberg sensor. For the electrically small antenna, length $L$ is

2 mm and resistance $R$ is 50 Ω, with resonant frequency tuned by adjusting inductance $L$. The Rydberg sensor consists of $10^4$ Rb atoms populating on the target state $|100D_{5/2,m_J=1/2}\rangle$. Colors in (a) and (b) correspond to consistent experimental conditions.

To demonstrate the optimal communication performance of the system, we performed parameter feedback cooling and pumped to high vacuums of $2\times10^{-7}$ mbar while tuning the resonant frequency by adjusting the trapping optical power. With a transmission rate set at 0.5 kbit/s, the experimental results shown in Fig.3. By adjusting the optical power, we achieved resonant frequencies from 107kHz to 142kHz along the $x$ axis, 149kHz to 176kHz along the $y$ axis, and 38kHz to 45kHz along the $z$ axis. As shown in Fig.3(a), when the resonant frequency is in the typical range from around 100 kHz to 150 kHz, the minimum detectable electric field strength that ensures a BER below 0.1% is between 0.4V/m and 0.8V/m. Fig.3(b) demonstrates the minimum detectable electric field strength in one second determined by noises when the net charge is $202\pm2e$. The electric field detection sensitivities are better than $10\mu V/cm/Hz^{1/2}$ in a wide frequency range from 30kHz to 180kHz. Despite variations in minimum detectable electric field strength across different directions, with the $x$-axis and $y$-axis performing significantly better than the $z$-axis, the disparity remains within an order of magnitude. This indicates that the LPMR could be effectively developed as an omnidirectional receiving antenna with the capacity for flexible frequency tuning.

**Application.** Fig.4 presents the experimental outcomes of transmitting both black and white and color images, utilizing format encoding. These experiments were performed at $2\times10^{-7}$ mbar using the y-axis mode with the resonant frequency of $f_0 = 149$ kHz. The signal transmission rate is set as 0.5kbit/s and the carrier frequencies are selected as $f_1=151$ kHz and $f_2=147$ kHz around the resonant frequency. Decoding the transmitted images from the collected data revealed that the BERs increase inversely with signal amplitude, underscoring the method's effectiveness.

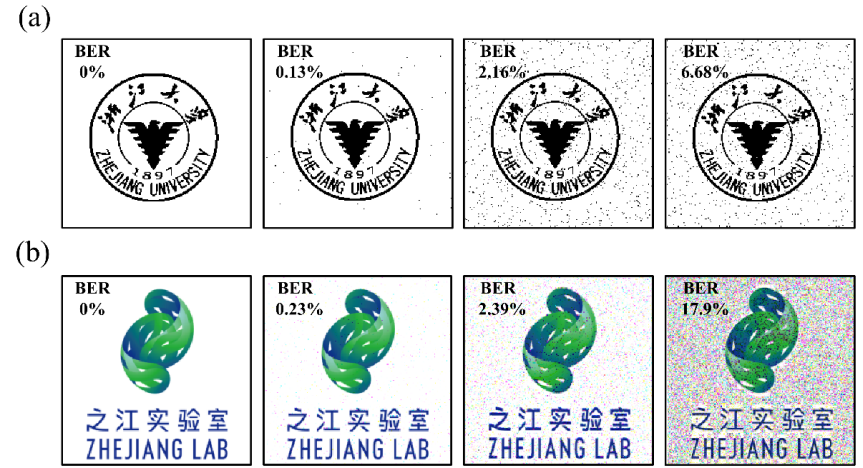

**Fig. 4.** Decoding results of image transmission. By changing the signal amplitude, decoded images with different BERs can be obtained. (a) Received and decoded images of *Zhejiang University* logo obtained with different BERs. (b) Received and decoded images of *Zhejiang Lab* logo obtained with different BERs. The former is a binary image with a pixel size of 200×200. Each pixel is encoded by 1 bit, with a total transmission time of 80s. The latter is an RGB image with a pixel size of 200×200. Each pixel is encoded by 24 bits, with a total transmission time of 1920s. BER is marked in the upper left corner of each image.

**Discussion.** Comparisons with the electric small antennas and Rydberg-atom-based sensors [59,60,61,62] are shown in Fig.3(b) as well. In contrast to the 2mm-long electric small antenna with comparable electric field detection sensitivity of the order of $1\mu V/cm/Hz^{1/2}$, the LPMR reduces the antenna size by nearly 4 orders of magnitude. Additionally, tuning the resonant frequencies from 40kHz to 170kHz requires adjusting the electrically small antenna's inductance from

29H to 408H, a challenging feat in practice. While our proposed nano-antenna offers miniaturization and easy frequency tuning, it cannot yet replace ~1m long commonly-applied electrically small antennas due to its 3-4 orders lower sensitivity. The sensitivity shows promising potential for improvement through furtherly enhancing the force detection sensitivity and increasing the net charge. Rydberg quantum sensors can serve as similar ultrasensitive receiving antennas, which are also research hotspots in electric field sensing. However, they are mainly applied for microwave and terahertz electric detection and do not possess advantages on both the bandwidth and the sensitivity in the LF communication. Moreover, Rydberg atom-based quantum sensors require multiple laser beams for pump and probe operation, whereas our method uses a single high-convergence laser beam for trapping and motion signal detection, simplifying the systems.

A significant advancement of LPMR is their independence from the size-frequency limitations of traditional antennas. The key metrics of different systems for LF communications are compared in Fig.5. In Fig. 5, LPMRs are shown to be at least four orders smaller in size compared to other antennas. Unlike conventional dipole antennas, LPMRs sensitively respond to electric fields through the exertion of electric forces. In this way, sensitive frequencies only depend on parameters of the trapping potential well and the nanoparticles. Consequently, by adjusting the particle's size, geometry, material, and the stiffness of potential wells, resonant frequencies can be tuned from the order of 10Hz to 100kHz. In addition, LPMRs oscillate in multiple orthogonal directions with distinct frequencies, enabling their application in frequency-division-multiplexing LF communications, another clear advantage.

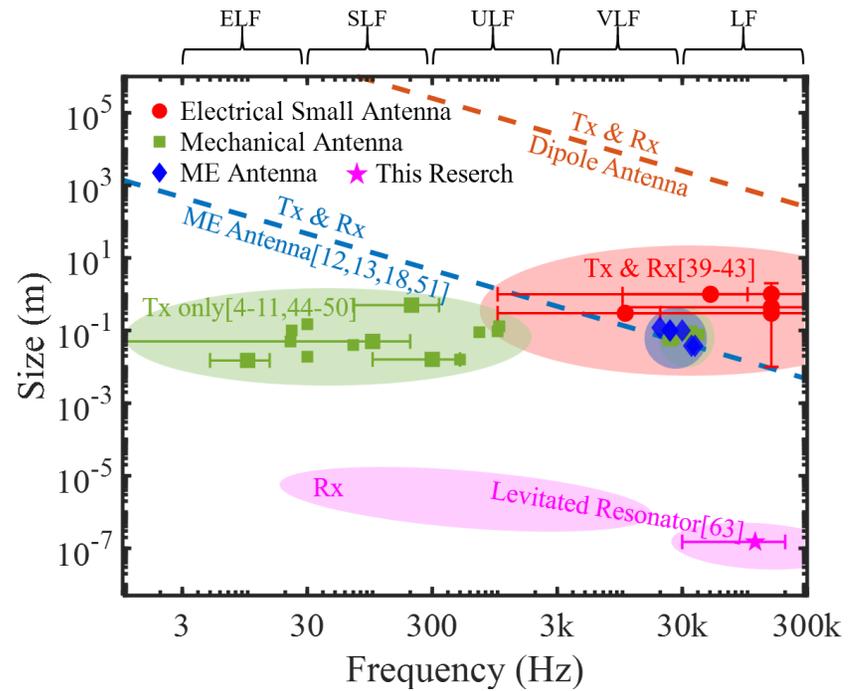

**Fig. 5.** Comparison of different resonant antennas vary in size and operational frequency for LF communications. The four types of antennas are compiled based on their transmission (Tx) and reception (Rx) capabilities: electrical small antennas[39-43], mechanical antennas [4-11,44-50], magnetoelectric antennas [12,13,18,51], and LPMRs. These antennas essentially operate as harmonic oscillators in response to electromagnetic signals. Dipole antennas, capable of both Tx and Rx, have sizes inversely proportional to their frequencies. Magnetoelectric antennas, constrained similarly in size-frequency, leverage shorter mechanical wave wavelengths to achieve size reductions by approximately four orders of magnitude. In contrast, traditional mechanical antennas, limited to transmission, do not follow a strict size-to-frequency correlation. Our LPMRs, designed as receivers, transcend this size-frequency limitation, diminishing low-frequency antenna sizes by nearly five orders of magnitude. While focusing on optically levitated nano-resonators, this method is extendable to various levitated micro- and nano-resonators [63] using optical, magnetic, and electrical traps, potentially lowering resonant frequencies further.

Although the bandwidth of a single nanoparticle in high vacuum degree is still relatively limited for the contradiction between the transmission rate and the receiving sensitivity for the 2FSK modulation. However, it is not a technical problem to expand the bandwidth of the system by constructing multiple potential wells [37,38], and the multiple potential wells obviously do not significantly increase the antenna size, so that our method can still play the advantage of miniaturization. In the future, we will also use the coulomb interaction between double particles to further improve the communication capability of the system.

**Conclusion.** In conclusion, we have presented a novel method for ultra-sensitive low-frequency (LF) communication using levitated nanoparticles as antennas, overcoming the traditional size-sensitivity limitations of conventional receivers. Utilizing the frequency shift keying (2FSK) modulation scheme, we established and validated a theoretical model linking error rates to transmission rates, signal strength, and vacuum degree. Further, using the high vacuum environment and the high charge state, we achieved superior LF communication performance with signal strength of approximately 0.1V/m and a bit error rate below 0.1% in tunable LF bands. Notably, our method drastically reduces the size of LF antennas from centimeters to the hundred-nanometer scale, a reduction of nearly five orders of magnitude, offering an effective solution for long-distance LF communication.

**Methods**

**Optical trap.** Our experimental setup is consistent with that described in refs [24,33,54]. A nanoparticle with net charge is optically levitated in a trap formed by focused laser beams at $\lambda$=1064nm. The measured diameter of silica particle (Nanocym-150) is $d$=142.8±3.3nm, with mass m= (3.06±0.26) ×$10^{-18}$kg. Motion information from the scattered light is detected by a four-quadrant photodetector. To ensure stable levitation in high vacuum conditions (P <$10^{-3}$mbar) and reduce the frequency drift caused by nonlinear effects, three-dimensional parametric feedback cooling of the center-of-mass motion of the particles is implemented. Specifically, the beam is modulated by an electro-optic modulator (EOM) before the objective lens. Compared to acousto-optic modulators (AOM), the use of EOM avoids phase delay on the order of hundreds of nanoseconds, which is beneficial for improving the feedback cooling effect.

**Communication module.** The optical trap is situated inside a metal vacuum chamber that reflects or attenuates LF electromagnetic interference outside the chamber. To generate a modulated electric field near the LPMR, a 2FSK digital communication signal generator is connected to a pair of needle electrodes located near the trap. The electrodes are placed horizontally perpendicular to the beam, with a diameter of 1mm and a separation of 2.52mm. One electrode is grounded, while the other is connected to a signal generator (Keysight 33500B) outside the vacuum chamber via a flange. This module allows adjustments of the carrier frequencies, symbol transmission rates, and signal amplitudes. For signal amplitudes over 10Vpp, a high-voltage amplifier (Aigtek ATA-2031) amplifies the signal before its connection to the electrode.

**Measurements of the net charge and the electric field.** Prior to conducting communication experiments, measuring the net charge and the electric field is essential. The method for measuring net charge and electric field follows the harmonically driven model detailed in reference [33,54]. For weak electric fields driving, the LPMR's displacement response is proportional to the electric field intensity at its location and its carried net charge. This response incrementally changes with each additional unit of charge. In practice, a single-frequency, nearly-resonant AC electric field is applied for net charge measurement. The charge is determined by dividing the response by the minimum step of response corresponds to a single charge. Then, the electric field measurement is based on the known net charge. The normalized measured values of three orthogonal components are $E_x$=202.08±5.69V/m, $E_y$=273.33±6.03V/m, $E_z$=65.07±1.24V/m, respectively. They equivalent to a 1V driving voltage at the electrodes, served as reference parameters. These parameters are then used to calculate the signal amplitude for analyzing communication BER, by multiplying with the electrode-applied voltage.

**Controlling the net charge.** Charge control in particles is commonly achieved through high-voltage ionization-induced plasma [30,52,54,55] or ultraviolet exposure-induced photoelectric effects [53,56], facilitating random control and precise charge measurement. However, achieving directed net charge control with stability poses a technical challenge. To overcome this, we utilize an electron beam method, where focused electron beam bombardment imparts a substantial negative charge on levitated particle surfaces, a technique proven effective in dusty physics [57, 58].

To enhance the application of electron beams in the field of levitation, it is crucial to optimally design the electron beam's energy, current, and spot size for stable particle levitation in high vacuum. The electron energy is partly used to overcome the Coulomb potential of the particle and partially converted into kinetic energy. Excessive energy raises the risk of the particle escaping the trap, while insufficient energy prevents free electrons from overcoming the Coulomb barrier, impeding their absorption onto the particle. For a levitated particle already with high net charge $Q=nq_e$, based on the point-charge model, the Coulomb potential that the electron needs to overcome to reach its final distance $r_{en}$ is given by:

$$E_{Cou}(n) = \frac{nq_e}{4\pi\varepsilon_0}\left(\frac{1}{r_{en}} - \frac{1}{r_{in}}\right) \quad (2)$$

Where $\varepsilon_0$ and $r_{in}$ represent the permittivity in vacuum and initial distance between the emitted electron and the particle, respectively. The required electron energy increases linearly with an increase in particle's charge. Additionally, the absorbed kinetic energy should be lower than the optical potential. For a common single-beam trap with $P$=100mW, the potential is approximately 90eV. Assuming a final distance of $r_{en}$=1nm, the required electron range for adjusting the charge from 10$e$ to 200$e$ is approximately 250eV.

Fig.5a illustrates a custom-made thermionic cathode electron gun positioned above the optical trap. It employs a thorium tungsten filament for electron emission. This setup allows for electron beam spot size adjustments from 2 to 10 mm and electron energy alterations ranging from 50 to 1000 eV, catering to charge control needs. The beam current is modifiable from 1nA to 400μA. The electron gun, mounted on a flange above the optical trap, measures about 10cm in length and 3.8cm in width, with a working distance of 10-50mm. The emitted electrons are focused and accelerated through multiple electrodes, and then guided by the anode to form an electron beam that irradiates the levitated particles. As shown in Fig.5b, the particle's net charge was successfully increased from 2$e$ to 202±2$e$ at 10$^{-6}$ mbar, surpassing the efficiency of traditional methods like high-voltage discharge and ultraviolet-induced photoelectron effects by an order of magnitude.

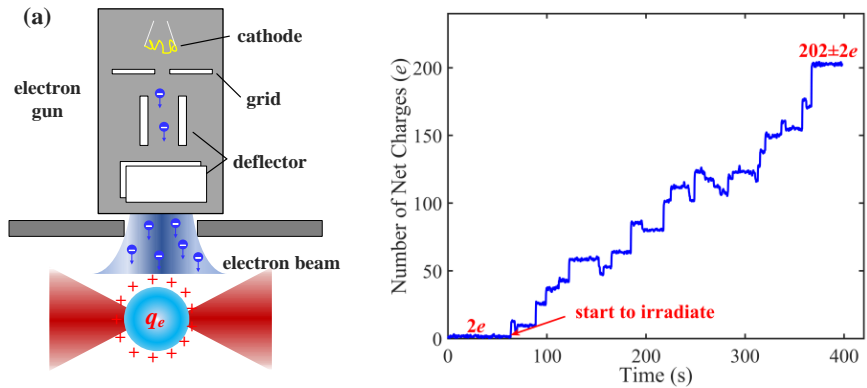

**Fig. 5.** Controlling the net charge. **a.** The thermionic cathode electron gun generates a focused electron beam to alter the charge to the negative direction. It comprises a barrel, cathode filament, grid, deflection electrode and anode. **b.** The initial charge of the particle at $10^{-6}$ mbar was $2e$ and then increased to $202\pm2e$ after a continuous 7-minute process of electron beam irradiation.

**Data availability.** All relevant data are available from the corresponding author upon request.

**Acknowledgements.** This research was supported by National Key Research and Development Program of China (2022YFB3203402), Major Project of Natural Science Foundation of Zhejiang Province (LD22F050002), Major Scientific Research Project of Zhejiang Lab (2019MB0AD01), institute-initiated Research Project of Zhejiang Lab (2022MB0AL02) and National Natural Science Foundation of China (62005248).

**Author contributions**

Z.F., X.G., C.L., Y.D. and H.H. initiated the original idea and designed the experiments. J.X., C.H. and Z.F. derived the theoretical and numerical model. P.H., J.X., X.Z. and N.L. build the optical platform. Z.C. and Y.Z. designed and assemble mechanical and vacuum structures. X.Z. and X.C. helped with the electronic equipment. J.X., S.Z. and C.H. performed the experiments. J.X., Z.F., S.Z. and C.H. analyzed the data., Z.F., J.X., S.Z., C.H. and H.C. discussed the results. Z.F., C.H., S.Z. and S.Z. wrote the manuscript. Z.F. and X.G. supervised the project.

**Disclosures.** The authors declare no conflicts of interest.

## Appendix A: principle and processes of communication demodulation

The optically levitated particle mechanical resonators (LPMRs) inherently oscillate within their traps at frequencies ranging from tens to hundreds of kilohertz. Its intrinsic oscillation is modulated by electric field signals to be transmitted, making it an excellent candidate for a LF receiving antennas. Modulation is facilitated by pre-controlling the net charge and applying electric-field driving forces. The LPMRs' modulated oscillational signal is acquired in the time domain by a data acquisition system (DAQ), and digital signal processing (DSP) is used for demodulation. The communication process's bit error rate (BER) is then determined by comparing the demodulated symbols with the original transmitted data. In the following, we explain the communication principle and complete processes of the LF demodulation in detail.

**(1) Intrinsic oscillations of the LPMRs**

Considering the thermal noise only, the intrinsic oscillations $x(t)$ is expressed by the Langevin equation [1]:

$$x''(t) + \Gamma_0 x'(t) + \omega_0^2 x(t) = F_{th}(t)/m \quad (A1)$$

where $\Gamma_0$, $\omega_0$, $m$, and $F_{th}(t)$ stand for the damping rate, the resonant frequency, the mass of the particle and thermal forces, respectively, and $F_{th}(t)$ is usually treated as a gaussian random process:

$$\langle F_{th}(t) \rangle = 0$$
$$\langle F_{th}(t) F_{th}(t') \rangle = 2mk_B T \Gamma_0 \delta(t-t') \quad (A2)$$

where $k_B$ and $T$ stand for the Boltzmann constant and temperature respectively. By performing a Fourier transformation $\tilde{x}(\omega) = \int x(t) e^{j\omega t} dt$, it is straightforward to get:

$$\tilde{x}(\omega) = \frac{\tilde{F}_{th}(\omega)}{m\left[(\omega_0^2 - \omega^2) + j\omega\Gamma_0\right]} \quad (A3)$$

where $\tilde{F}_{th}(\omega) = \int F_{th}(t) e^{j\omega t} dt$.

**(2) 2FSK modulation of the transmitted signal**

The transmitted signal $s(t)$ consists of binary signals $u_1(t) = V_0 \cos(2\pi f_1 t)$ and $u_2(t) = V_0 \cos(2\pi f_2 t)$, representing bits "1" and "0", respectively. And a typical 2FSK signal $s(t)$ to be transmitted writes:

$$s(t) = S \cdot V_0 \cos(2\pi f_1 t) + (1-S) \cdot V_0 \cos(2\pi f_2 t) \quad (A4)$$

where $S=[b_1 b_2 \cdots b_n] \cdot t_s$ represents original symbol with $[b_1 b_2 \cdots b_n]$ a sequence of bits "1" and "0", and $t_s$ stands for the duration of a single bit determined by the transmission rate $r_s$, $t_s = 1\text{bit}/r_s$.

The signal $s(t)$ exerts electric-field driving forces $F_{dr}(t)$ on the particle:

$$F_{dr}(t) = Nq_e E_0 s(t) \quad (A5)$$

where $q_e$, $N$ and $E_0$ (unit in m$^{-1}$) stand for the elementary charge, the charge number carried by the nanoparticle, and the normalized electric-field intensity [2], respectively and thus $F_{dr}(t)$ can be regarded as an AC time-domain signal with an amplitude of $F_0 = Nq_e E_0 V_0$. Adding electric-field driving forces $F_{dr}(t)$ to the Langevin equation (A1), the motional signal $\tilde{x}_s(\omega)$ in frequency-domain can be obtained:

$$\tilde{x}_s(\omega) = \frac{\tilde{F}_{th}(\omega) + \tilde{F}_{dr}(\omega)}{m\left[(\omega_0^2 - \omega^2) + j\omega\Gamma_0\right]} \quad (A6)$$

where $\tilde{F}_{dr}(\omega) = \int F_{dr}(t) e^{j\omega t} dt$ reflects the additive modulation on the particle's frequency-domain oscillational signal from the transmitted signal $s(t)$. This makes the particle a sensitive and nano-antenna in LF communication systems as $\tilde{x}_s(\omega)$ is detectable at the subsequent detection module.

**(3) Demodulation of the transmitted signal**

By detecting the time-domain signal with the DAQ, $\tilde{x}_s(\omega)$ is acquired through a Fourier transform. The subsequent demodulation process includes Fourier transform, transfer function compensation, frequency domain bandpass filtering, inverse Fourier transform, full-wave rectification, time domain lowpass filtering and symbol decision. The whole process is interpreted in detail as follows.

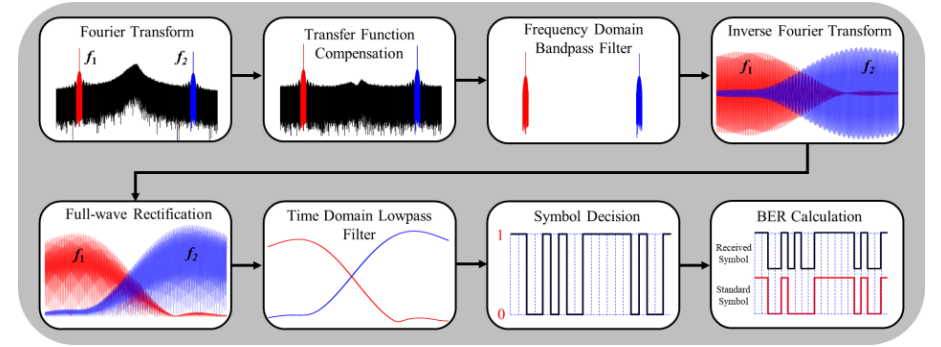

**Fig. S1.** The complete processes of the LF demodulation.

According to Eq. A6, the nanoparticle's modulation in response to electric-field driving forces varies with frequency. This frequency-dependent response is quantitatively analyzed by defining the transfer function of the LPMR:

$$\chi(\omega) = m^{-1}\left[(\omega_0^2 - \omega^2) + j\omega\Gamma_0\right]^{-1} \quad (A7)$$

Fig.S2a shows the frequency spectrum of the detected motional signals $\tilde{x}_s(\omega)$. The two carrier frequencies and intrinsic frequencies are 130kHz, 170kHz and 150kHz respectively, and $r_s$=1kbps. The transfer function causes differing amplitudes at the two carrier frequencies. And the transfer function compensation is addressed through a straightforward division process:

$$D(\omega) = \frac{\tilde{x}_s(\omega)}{\chi(\omega)} = \tilde{F}_{th}(\omega) + \tilde{F}_{dr}(\omega) \quad (A8)$$

After compensation, amplitudes at both carrier frequencies equalize (Fig.S2b). The desired signal exceeds noise near carrier frequencies, while noise dominates at distant regions. Hence, it is crucial to filter distant noise before recovering original symbol. In addition to noise suppression, filtering is essential for envelope demodulation. The goal of envelope demodulation is to extract the time-domain envelope of the detected motional signal, which should comprise pulses with a consistent duration of $t_s$. Envelope demodulation is performed on two carrier frequencies, resulting in the following two bandpass filters $H_{env}^{(1)}(\omega)$ and $H_{env}^{(2)}(\omega)$:

$$H_{env}^{(1)}(\omega) = \begin{cases} 1, f_1 - f_T < \dfrac{\omega}{2\pi} < f_1 + f_T \\ 0, others \end{cases}$$

$$H_{env}^{(2)}(\omega) = \begin{cases} 1, f_2 - f_T < \dfrac{\omega}{2\pi} < f_2 + f_T \\ 0, others \end{cases} \quad (A9)$$

where $f_T$ stands for the baseband signal's bandwidth and quantitatively equals to the transmission rate $r_s$. The filtered frequency spectrum is shown in Fig.S2c.

After filtering and inverse Fourier transforms, we can recover two electric-field driving force signals $F_{dr}^{(1)}(t)$ and $F_{dr}^{(2)}(t)$ at two carrier frequencies in the time domain:

$$F_{dr}^{(1)}(t) = \frac{1}{2\pi}\int D(\omega)H_{env}^{(1)}(\omega)e^{j\omega t}d\omega$$

$$F_{dr}^{(2)}(t) = \frac{1}{2\pi}\int D(\omega)H_{env}^{(2)}(\omega)e^{j\omega t}d\omega \quad (A10)$$

The envelope detections are obtained through the full-wave rectification and the following time-domain lowpass filtering. Full-wave rectification retains the value of time-domain signal only while throws its phase information away as is shown in Eq. A11. The cut-off frequency of the following time-domain lowpass filtering equals to the baseband signal's bandwidth $f_T$ and the final detected envelopes are shown in Eq. A12.

$$F_{dr,rect}^{(1)}(t) = \left|F_{dr}^{(1)}(t)\right|$$
$$F_{dr,rect}^{(2)}(t) = \left|F_{dr}^{(2)}(t)\right| \quad (A11)$$

$$u_r^{(1)}(t) = \frac{1}{2\pi}\int \tilde{F}_{dr,rect}^{(1)}(\omega)H_{LP}(\omega)e^{j\omega t}d\omega$$
$$u_r^{(2)}(t) = \frac{1}{2\pi}\int \tilde{F}_{dr,rect}^{(2)}(\omega)H_{LP}(\omega)e^{j\omega t}d\omega \quad (A12)$$

where $\tilde{F}_{dr,rect}^{(1,2)}(\omega) = \int F_{dr,rect}^{(1,2)}(t)e^{-j\omega t}dt$ and $H_{LP}(\omega)$ stands for the transfer function of corresponding lowpass filtering.

Through sampling, two sequences are obtained according to Eq. A13 and the sampling starts from $t = t_s/2$ with the interval of $t_s$.

$$S_r^{(1)} = \left[b_{r1}^{(1)}b_{r2}^{(1)}b_{r3}^{(1)}\cdots b_{rn}^{(1)}\right]\cdot t_s$$
$$S_r^{(2)} = \left[b_{r1}^{(2)}b_{r2}^{(2)}b_{r3}^{(2)}\cdots b_{rn}^{(2)}\right]\cdot t_s \quad (A13)$$

The final recovered symbol sequence $S_r = [b_{r1}b_{r2}b_{r3}\cdots b_{rn}]\cdot t_s$ is obtained by symbol decisions as follows:

$$b_{ri} = \begin{cases} 1, b_{ri}^{(1)} > b_{ri}^{(2)} \\ 0, b_{ri}^{(1)} < b_{ri}^{(2)} \end{cases} i = 1,2,3,\cdots,n \quad (A14)$$

Comparing the recovered symbol sequence $S_r = [b_{r1}b_{r2}b_{r3}\cdots b_{rn}]\cdot t_s$ to the original symbol sequence $S = [b_1 b_2 \cdots b_n]\cdot t_s$, the measured bit error rate (BER) is finally obtained as follows:

$$BER_{mea} = \frac{1}{n}\sum_{i=1}^{n}\left|b_{ri} - b_i\right| \quad (A15)$$

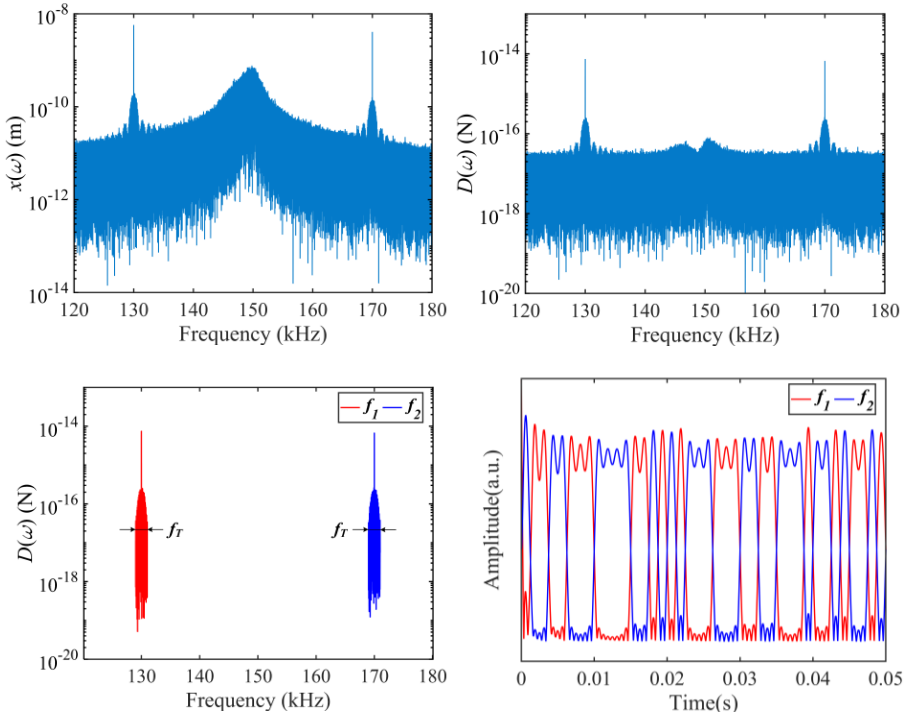

**Fig. S2. Demodulation of the transmitted signal. a.** Frequency spectrum of the detected motional signals. **b.** The compensated frequency spectrum. **c.** The band pass filtered frequency spectrum. **d.** The time-domain signals obtained by envelope detection, and the recovered symbol sequence can be derived by sampling and evaluation of these signals.

**Appendix B: theoretical model for deriving the BER**

In a typical 2FSK communication system with additive white Gaussian noise, the theoretical BER of the system is determined by the modulated signal amplitude $V_0$ [3]:

$$\text{BER} = \frac{1}{2}\exp(-r/2) \quad (B1)$$

where $r = V_0^2/2\sigma_n^2$ with $\sigma_n$ the standard deviation of the white noise.

In our communication system, the particle's intrinsic oscillation acts as colored noises, which restricts the application of Eq. B1 in evaluating the BER. Fortunately, as has been shown in Appendix A, by compensated with the transfer function $\chi(\omega)$, the demodulated signal $D(\omega)$ is a superposition of thermal noise and electric-field driving force signals in the frequency domain. The former is a Gaussian white noise with a one-sided PSD of $S_{th} = 4mk_BT\Gamma_0$, while the latter is the desired demodulated signal with a corresponding time-domain amplitude of $F_0$. In this way, Eq. B1 can be applied to evaluate the BER.

The filtering bandwidth $B$ equals twice the transmission rate $f_T$ (i.e., $B = 2f_T$) and the standard deviation of the Gaussian white noise is $\sigma_n = \sqrt{4mk_BT\Gamma_0 B} = \sqrt{8mk_BT\Gamma_0 f_T}$. Substituting is and the amplitude $F_0$ into Eq. B1, finally the BER in our system is derived:

$$\text{BER} = \frac{1}{2}\exp\left[-\frac{(E_V Nq_e)^2}{32k_BTm\Gamma_0 f_T}\right] \quad (B2)$$

Considering the thermal noise only, the BER is determined by the net charge $Nq_e$, the electric field signal amplitude $E_V = V_0E_0$. This equation serves as the fundamental basis for assessing and improving the performance of the communication system.

According to error transfer functions, the error of BER depends on errors of $E_V$, $m$, $T$, $\Gamma_0$ and $f_T$:

$$\sigma_{\mathrm{BER}} = \sqrt{\left(\frac{\partial}{\partial E_v}BER\right)^2 \sigma_{E_v}^2 + \left(\frac{\partial}{\partial m}BER\right)^2 \sigma_m^2 + \left(\frac{\partial}{\partial T}BER\right)^2 \sigma_T^2 + \left(\frac{\partial}{\partial \Gamma_0}BER\right)^2 \sigma_{\Gamma_0}^2 + \left(\frac{\partial}{\partial f_T}BER\right)^2 \sigma_{f_T}^2} \qquad (B3)$$

Substituting expressions of partial derivatives above, we can get the error of BER in Equation(B4).

$$\sigma_{BER} = BER \frac{E_V^2 N^2 q_e^2}{32 m k_B T \Gamma_0 f_T} \sqrt{\frac{4\sigma_{E_V}^2}{E_V^2} + \frac{\sigma_m^2}{m^2} + \frac{\sigma_T^2}{T^2} + \frac{\sigma_{\Gamma_0}^2}{\Gamma_0^2} + \frac{\sigma_{f_T}^2}{f_T^2}} \qquad (B4)$$